\documentclass[aps,amssymb,prl,showpacs,preprint ]{revtex4} \fussy

\usepackage{graphicx}
\usepackage{dcolumn}

\begin{document}

\newcommand{\twenteadres}{Complex Photonic
Systems, MESA$^+$ Institute for Nanotechnology and Department of
Science $\&$ Technology, University of Twente, P.O. Box 217, 7500
AE Enschede, The Netherlands}

\title{
Optical extinction due to intrinsic structural variations of
photonic crystals}

\author{A. Femius {Koenderink}}
\altaffiliation[Present address: ]{Nano-Optics Group, Laboratory
for Physical Chemistry, Swiss Federal Institute of Technology
(ETH), Z\"urich, Switzerland}  \affiliation{\twenteadres}

\author{Ad Lagendijk}
\altaffiliation[Also at: ]{FOM Institute for Atomic and Molecular
Physics AMOLF, Center for Nanophotonics, Kruislaan 407, 1098 SJ
Amsterdam, The Netherlands} \affiliation{\twenteadres}

\author{Willem L. Vos}

\altaffiliation[Also at: ]{FOM Institute for Atomic and Molecular
Physics AMOLF, Center for Nanophotonics, Kruislaan 407, 1098 SJ
Amsterdam, The Netherlands} \affiliation{\twenteadres}
\email{W.L.Vos@utwente.nl} \homepage{www.photonicbandgaps.com}

\date{Prepared for Phys. Rev. B, June 13th, 2005.}

\begin{abstract}
Unavoidable variations in size and position of the building blocks
of photonic crystals cause light scattering and extinction of
coherent beams. We present a new model for both 2 and
3-dimensional photonic crystals that relates the extinction length
to the magnitude of the variations. The predicted lengths agree
well with our new experiments on high-quality opals and inverse
opals, and with literature data analyzed by us. As a result,
control over photons is limited to distances up to 50 lattice
parameters ($\sim~15~\mu$m) in state-of-the-art structures,
thereby impeding large-scale applications such as integrated
circuits. Conversely, scattering in photonic crystals may lead to
novel physics such as Anderson localization and non-classical
diffusion.

\end{abstract}

\pacs{42.70.Qs, 42.25.Dd, 42.25.Fx, 81.05.Zx}

\maketitle

The promise of full control over emission and propagation of light
has led to a widespread pursuit of photonic crystals in recent
years~\cite{soukoulis01}. Photonic crystals are dielectric
structures in which the refractive index varies periodically over
length scales comparable to the wavelength of light. For
three-dimensional periodicities, such crystals promise a photonic
band gap, \emph{i.e.}, a frequency range for which emission and
propagation of light are completely forbidden. Ideally, photonic
band gap crystals will form a backbone in which many photonic
devices, such as ultrasmall waveguides, cavities and light
sources, are combined to create optical integrated
circuits~\cite{noda03}. This requires photonic crystals with
negligible optical extinction over millimeter
distances~\cite{noda03}.

Tremendous progress has been made in the fabrication of photonic
bandgap materials of the required high refractive
index-materials~\cite{wijnhoven,blanco00,noda00,vlasov01}, with
low point and plane defect densities~\cite{vlasov01}. Structural
variations in size and position of the building blocks, however,
are intrinsic to three- and two-dimensional (3D, resp. 2D)
photonic crystals alike, amounting to at least 2 to 7\% of the
lattice spacing in all current state-of-the-art photonic
crystals~\cite{wijnhoven,babacrete00}. While displacements are
well-known in condensed matter~\cite{ashcroft76}, size
polydispersity of individual unit cell building blocks, including
roughness, is intrinsic to meta-materials such as photonic
crystals. All such variations can ultimately be traced back to
basic thermodynamic arguments~\cite{ashcroft76}, but are at
present probably limited by materials science. These deviations
from perfect periodicity cause scattering, and hence exponential
attenuation of coherent beams propagating through photonic
crystals over lengths $\ell$, also known as the `(extinction) mean
free path'. After propagating over a distance $\ell$, a coherent
light beam is converted to a diffuse glow that corrupts the
functionality of any photonic integrated circuit. Conversely,
short mean free paths open up novel physics related to diffusion
of light and ultimately Anderson localization of
light~\cite{John87,sheng}. Therefore, it is crucial to obtain the
relation between the extinction length $\ell$ and the structural
disorder. In this paper, we derive such a relation and test it
against available experimental results \cite{Koenderink04}.
\begin{figure}[t]
\centerline{\includegraphics[width=0.9\columnwidth]{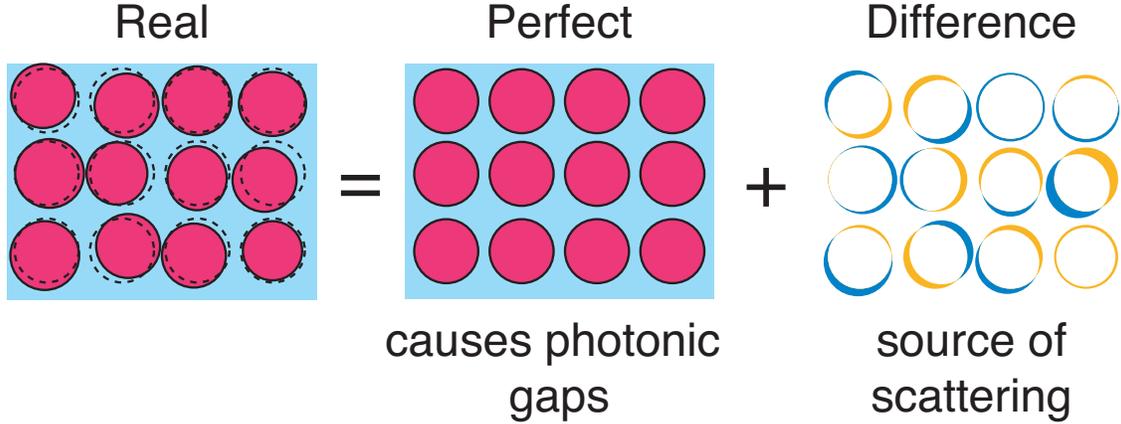}}
\caption{(Schematic) Any 2D or 3D real photonic crystal is an
ordered stack of building blocks with a spread $\Delta r$ in their
average radius $r$, each slightly displaced (displacement $\Delta
u$) from the  lattice sites.  The real structure is the sum of the
perfect crystal and the difference between the real and perfect
structure. This difference is a collection of thin shells that
each scatter weakly. Due to their high number density, the shells
dominate the  scattering loss. \label{fig:scattmech}}
\end{figure}

We consider extinction in photonic crystals due to scattering by
size polydispersity and displacements from lattice sites of the
structural units (size $r$) that compose the unit cell (lattice
spacing $a$). Light scattering is caused only by the
\emph{difference\/} in refractive index profile of the displaced,
slightly polydisperse building blocks as compared to the ideally
ordered structure. As illustrated in Fig.~\ref{fig:scattmech},
this difference is a collection of thin shells of high and low
index material. The polydispersity and displacements of the
building blocks translate linearly into the shell thickness
$\Delta r$. Since in many photonic crystals, such as cubic (3D) or
hexagonal (2D) structures,  light transport is isotropic, we treat
the ideal crystal as an effectively homogeneous medium with index
$n_{\rm eff}$ equal to the volume-averaged refractive
index~\cite{averagenote}. Within this framework, the inverse
extinction length
\begin{equation} \frac{1}{\ell}= \rho\cdot \sigma_{\rm Rayleigh}
\cdot F \label{eq:scattbase}\end{equation}
is the product of three factors ~\cite{hulst81}: Rayleigh's
extinction cross section $\sigma_{\rm Rayleigh}$ of each shell,
the number density of shells $\rho$, and a wavelength-dependent
geometrical factor $F$ which embodies corrections beyond Rayleigh
scattering~\cite{transportnote}. Since the volume of each shell is
proportional to its thickness $\Delta r$, Rayleigh's extinction
cross section is proportional to $(m-1)^2 \Delta r^2$, where $m$
is the  index contrast relative to the background medium. Even
though scattering by each shell is generally weak, the huge
density $\rho$ set by the number of structural units per unit cell
causes the scattering mechanism to be important. For Rayleigh
scatterers, in the low-frequency limit, the dimensionless factor
$F$ equals unity.  For weakly scattering shells, the Rayleigh-Gans
approach is suited to find $F$~\cite{hulst81,transportnote}.
\begin{figure*}[th]
\centerline{\includegraphics[width=0.8\textwidth]{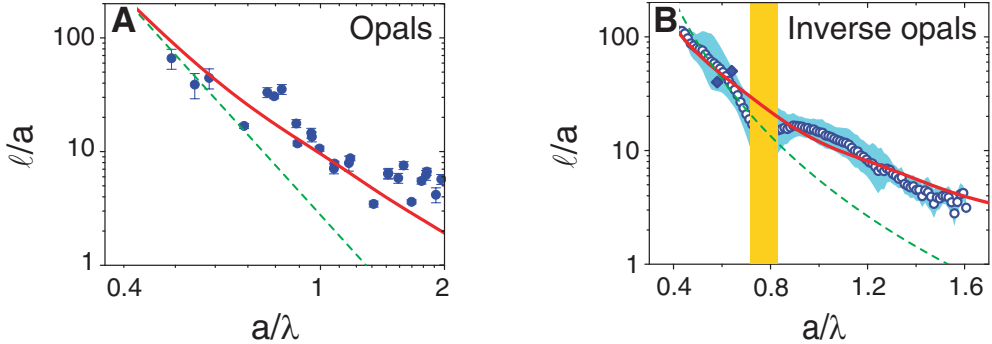}}
\caption{(color) Symbols: mean free path~\cite{transportnote}
$\ell$ in units of $a$ versus normalized frequency $a/\lambda$ in
polystyrene opals (A) and titania inverse opals (B). Open symbols
in (B) were obtained by averaging for each $a/\lambda$ total
transmission spectra for many samples with different $a$. The blue
shaded area indicates the standard deviation. In the stop gap
(orange bar), total transmission is reduced in excess of $\ell/L$
due to Bragg reflection of the input beam. This affects the data
in this limited range. Previous data show that $\ell$ is
unaffected if the frequency is tuned through a
gap~\cite{koenderink00}. In both (A) and (B), the extinction
length agrees well with the model
(\ref{eq:scattbase},\ref{eq:shellmfp}) with $\delta R \sim 5\%$
(red curves). Green curves  represent scaling of $\ell$ with
$\omega^{-4}$, and illustrate the failure of Rayleigh scattering
models.\label{fig:ourmfp}}
\end{figure*}

We now focus on the extinction length in 3D photonic crystals that
consist of spheres (mean radius $r$), such as opals and inverse
opals where many data are available. Size polydispersity results
in scattering due to thin spherical shells with a Gaussian
distribution of thicknesses. The inverse extinction length $\ell$
scales quadratically with the size polydispersity $\Delta r$ and
with $m-1$, since Rayleigh's extinction cross section for a shell
of thickness $\Delta r$ reads $\sigma_{\rm Rayleigh}=(32\pi/3)
(m-1)^2 k_{\rm eff}^4r^4 \Delta r^2$ (cf.~\cite{hulst81}), with
$k_{\rm eff}$ the wave vector in the effective medium. We find
that the Rayleigh-Gans correction~\cite{hulst81}
\begin{equation}F(k_{\rm eff} r) \approx 0.78 \frac{1}{(k_{\rm eff}
r)^2}(1+0.09 k_{\rm eff} r), \label{eq:shellmfp}\end{equation}
reduces the well-known fourth power Rayleigh increase of
extinction to a nearly quadratic dependence on wave vector
~\cite{approxnote}. We have checked the validity of our result
using the exact Mie-solution for spherical shells. Although for
$m>2$ and $k_{\rm eff} r >1$ the Rayleigh-Gans result
underestimates the extinction loss compared to Mie theory, the
Mie-model reproduces the quadratic scaling with frequency and
shell thickness. Our model captures both the effect of
polydispersity $\Delta r/r$ and displacements $\Delta u/r$:
calculations of $F$ show that both effects are similar in
magnitude, and can be combined by taking an \emph{effective\/}
shell thickness $\Delta r+0.5\Delta u$. From now on, $\delta R$
indicates \emph{effective\/} shell thicknesses normalized by the
shell radius. An essential result of our paper is that given the
current fabrication accuracies of $\delta R \sim 5\%$, the maximum
extinction length $\ell$ is only 50 lattice spacings in high-index
crystals at relevant frequencies.

Enhanced backscattering measurements obtained earlier by us have
allowed us to determine the mean free path
$\ell$~\cite{transportnote} in synthetic opals, \emph{i.e.}, fcc
crystals of close packed polystyrene spheres with $n=1.59$ and
$n_{\rm eff}=1.45$~\cite{koenderink00}. In
Figure~\ref{fig:ourmfp}(A), we plot $\ell$ for a wide normalized
frequency range, obtained with $\lambda=632, 685$ and $780$~nm,
and many different $a$. We see that $\ell$ decreases from $100a$
for frequencies below first order diffraction, to $5a$ at the
highest frequencies, where we have converted the wavevector from
the scattering model to the frequency scale $a/\lambda$ typical of
photonic crystals. The data and our model agree well on both the
observed decrease of $\ell$ with $a/\lambda$ and the magnitude of
$\ell$, which confirms that extinction is due to non-uniformities
and displacements of the spheres, assuming $\delta R = 5\%$. This
value matches well with the cumulative effect of polydispersity
$\sim 2\%$ and rms displacements of spheres from their lattice
sites ($\leq 3.5\%$ of the nearest neighbor distance), as
independently determined by small angle X-ray
scattering~\cite{megens01}. In contrast, the data refute the often
assumed Rayleigh $\omega^4$ dependence~\cite{vlasov99b,blanco00}.
The degree of extinction is also inconsistent with the common
assumption that scattering is due to point defects, \emph{e.g.\/},
missing spheres: From the cross-section of a sphere~\cite{hulst81}
we calculate that the observed scattering would require a density
of missing spheres larger than $0.13a^{-3}$, an order of magnitude
larger than the estimated density
$0.01a^{-3}$~\cite{vlasov99b,vlasov01}.

We have carried out new experiments to probe scattering losses in
photonic crystals with high photonic interaction strength, i.e.
inverse opals in a TiO$_2$ backbone. The strength of the
interaction of a photonic crystal with light is gauged by the
relative bandwidth $S$ of the lowest order gap in the dispersion
relation, see Ref.~\cite{soukoulis01}, p. 194. The generally
pursued large interaction strengths require a large index contrast
$n_{\rm high}/n_{\rm low}$ and are thus associated with stronger
scattering, due to the factor $(m-1)^2$ in Rayleigh's cross
section. While the magnitude of the non-uniformities is similar to
those in the direct opals~\cite{wijnhoven}, the inverse opals
present a much larger index contrast $n=2.7\pm 0.4$ ($n_{\rm
eff}\approx 1.18$). We have determined the frequency dependence of
$\ell$ from total diffuse transmission ($T=\ell/L$, with $L$ the
sample thickness~\cite{transportnote}). We used white-light FTIR
spectroscopy to cover a wide normalized frequency range for many
samples with $a=650$~nm to $930$~nm. To obtain the absolute
magnitudes of the mean free paths we calibrated our measurements
by measuring the absolute values of the transmission (closed
symbols) and using enhanced-backscattering
data~\cite{koenderink00}. Figure~\ref{fig:ourmfp}(B) shows that
$\ell$ decreases from $100a$ at $a/\lambda=0.4$ to only $4a$ at
$a/\lambda=1.6$. This decrease of $\ell$ for the inverse opals is
in excellent correspondence with our prediction (solid curve),
taking a non-uniformity $\delta R =4\%$ that is consistent with
independent structural data~\cite{wijnhoven}.
\begingroup
\squeezetable
\begin{table}[b]
\caption{Photonic interaction strength $S$, structure and
extinction in 3D photonic crystals.\label{tab:fitvalues}}
\begin{tabular}{l|cccccc}
\hline \hline Ref.      & $S$ & $n_{\rm sphere}$ / $n_{\rm
inter}$\footnotemark[1]
      &
      $r/a$
      & $\ell/a$\footnotemark[2] & {$x_{\rm exp}(x_{RG})$}\footnotemark[3] & $\delta R$\footnotemark[4]\\
\hline  \cite{pradhan97}\footnotemark[5]
      &  $0.7\% $  & $1.59   / 1.33$       & $0.116$\footnotemark[6]& $10^5$    & $4(3.3)$  & $12\%$\\
 \cite{koerdt03}\footnotemark[5]       & $< 1\%$    & $1.42 / 1.48 $           & cp\footnotemark[7]        & $3000$             &  $3(2.6)$  & $6\%$   \\  
 \cite{tarhan96}\footnotemark[5]       & $1.6\%$    & $1.59  / 1.33$          & $0.143$   & $1000$             &  $3.3(3)$  & $15\%$   \\  
 \cite{vlasov99b}\footnotemark[5]      & $2\% $     & $1.32  / 1.47 $      & cp        & $700$              &  $2.6(2.6)$  & $6\%$   \\  
 \cite{park99}\footnotemark[5]         & $3\%$      & $1.59  / 1.33   $        & cp        & $100$              &  $\leq 2(2.5)$   & $7\%$    \\  
 \cite{huang01}\footnotemark[8]        & $5\%$      & $1.41  / 1.0    $         & cp        & $17$               &  $\cdots$         & $9\%$    \\
 \cite{miguez97}\footnotemark[5]\footnotemark[9]       & $5.5\% $   & $1.45 /1.0   $           & cp        &  $\cdots$               & $2.6(2.4)$    &$\cdots$     \\ 
Fig.~\ref{fig:ourmfp}(a)\footnotemark[8]     & $7\%$    & $1.59  /  1.0  $         & cp        & $  50 $            & $1.8(2.4)$    & $ 5\%$   \\
Fig.~\ref{fig:ourmfp}(b)\footnotemark[9]  & $11\%$      & $1.0   / 2.7$             & cp        & $ 40 $             & $2.6(2.5)$    & $4\%$    \\
\hline \hline
 \end{tabular}

\footnotemark[1]{$n_{\rm sphere}/n_{\rm inter}$: refractive
indices of spheres, resp. background medium.} \footnotemark[2]{The
$\ell/a$ are for $a/\lambda$ in the first stop gap.}
\footnotemark[3]{Decay powers $x$ obtained by fitting $\ell\propto
\omega^{-x}$ to  the data ($x_{\rm exp}$), resp. model
Eq.~(\ref{eq:scattbase},\ref{eq:shellmfp}) in the same frequency
range ($x_{RG}$).} \footnotemark[4]{Effective shell radii that
best fit the data over the full available frequency range.}
\footnotemark[5]{Transmission}. \footnotemark[6]{Bcc instead of
fcc}. \footnotemark[7]{cp=close-packing, $r/a=1/\sqrt{8}$.}
\footnotemark[8]{Enhanced backscattering}.
\footnotemark[9]{Diffuse total transmission.}
\end{table}
\endgroup

To further test the validity of our model, we have analyzed
transmission data reported in many papers encompassing fcc and bcc
photonic crystals, with sphere volume fractions from
$\varphi=0.7\%$ to $74\%$ and index contrasts $n_{\rm high}/n_{\rm
low}$ from $1.05$ to
$1.5$~\cite{vlasov99b,tarhan96,koerdt03,pradhan97,miguez97,park99,huang01}.
Extinction causes the coherent-beam transmission outside stop gaps
to decrease according to Lambert-Beer's law $T_{coh}=e^{-L/\ell}$.
In all cases, except of course for the dilute
crystal~\cite{pradhan97}, fitting a power law dependence
$\ell\propto\omega^{-x}$ to each data set shows that extinction
does not increase according to Rayleigh's law. Indeed,
Table~\ref{tab:fitvalues} shows that we find exponents $x<4$ in
reasonable correspondence to the exponents predicted by our model
in the same frequency windows. Similar exponents were recently
also observed in Ref.~\cite{rengarajan05}. Fits to our model
further show that extinction lengths for the wide range of
crystals agree with {$\delta R > 4\%$}, consistent with typical
sphere polydispersities and displacements of $2$--$5\%$, as
reported in Table~\ref{tab:fitvalues}. The quantitative agreement
of $\ell$ with Eqs.~(\ref{eq:scattbase},\ref{eq:shellmfp})
confirms that polydispersity and displacements of unit cell
building blocks determine  scattering loss in 3D photonic
crystals.

\begin{figure}[b]
\centerline{\includegraphics[width=0.75\columnwidth]{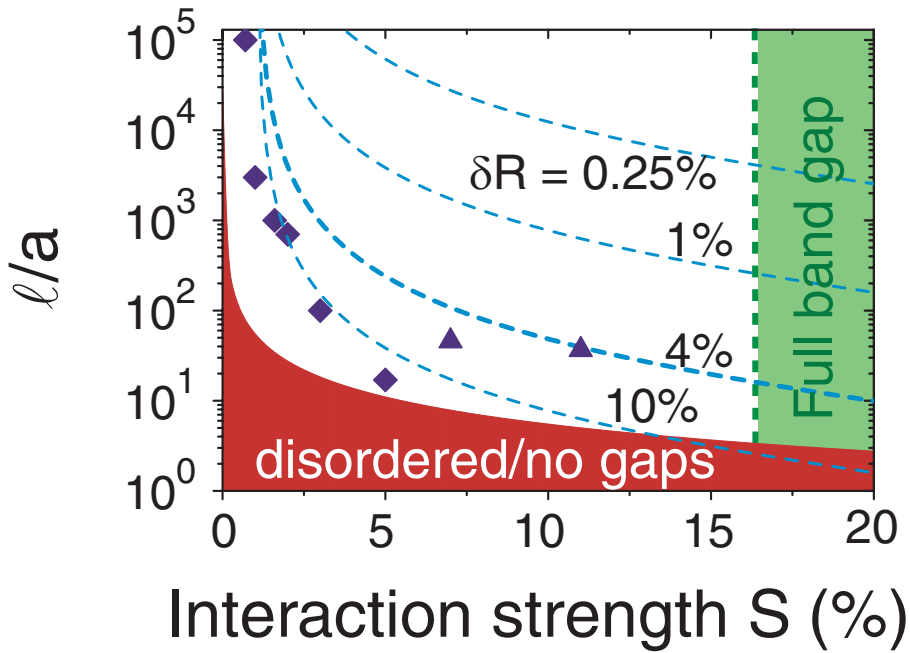}}
\caption{(color) Universal dependence of extinction length on
interaction strength $S$ in photonic crystals. Dashed curves:
extinction calculated for fcc inverse opals (assuming 26\% high
index material). Symbols: our data ($\blacktriangle$), and
literature data analyzed by us ($\blacklozenge$). Observed losses
are consistent with {$\delta R > 4\%$}. If $\ell$ is shorter than
the length needed for Bragg diffraction, structures are
essentially disordered (shaded red). Complete band gaps are
expected for $S>15\%$ (shaded green). Photonic crystal integrated
circuits require $\ell/a \geq 10^{4}$ at $S > 15 \%$, beyond
current state-of-the-art. \label{fig:Psivsell}}
\end{figure}
Given the success of our model, we can now use it to infer the
general dependence of the extinction length $\ell/a$ on the
photonic interaction strength $S$ and the non-uniformity $\delta
R$. In Figure~\ref{fig:Psivsell} we present both $\ell/a$ and $S$
that are calculated as a function of index contrast $(m-1)$. It is
clear that the extinction length decreases both with increasing
photonic strength and with increasing structural disorder. We also
present the experimental extinction data from
Table~\ref{tab:fitvalues} for fcc opals and inverse opals, showing
again a good agreement with our model with {$\delta R > 4\%$}. A
photonic band gap requires interaction strengths beyond $S=0.15$;
extinction lengths less than 20 lattice spacings are expected at
the current level of fabrication accuracy. Ultimately, one hopes
to realize photonic crystals that combine many optical functions.
Recent technology roadmaps foresee crystals containing $\sim 10^4$
optical functions per mm$^{2}$~(Ref. \cite{noda03}, p. 245),
requiring negligible loss over more than mm distances. From the
general scaling of extinction with non-uniformity we conclude that
applications of photonic band gap crystals in circuits require a
formidable tenfold increased perfection in statistical fabrication
accuracy to $\delta R < 0.25\%$, or subnanometer precision. Such
an improvement is far beyond the current
state-of-the-art~\cite{soukoulis01,noda03}.

Although 3D photonic crystals potentially offer the best platform
for photonic crystal functionality, 2D photonic crystals possess
many of the desired properties with the advantage of ease of
fabrication. While the fabrication methods are radically
different, 2D photonic crystals suffer from similar polydispersity
and displacements of their unit cell building blocks as 3D
crystals~\cite{babacrete00}. To obtain the scattering losses, we
consider 2D crystals of infinitely long cylinders. Now, Rayleigh's
cross section per unit length $\sigma_{\rm Rayleigh}
=(3\pi^2)(m-1)^2 k_{\rm eff}^3 r^2 \Delta r^2$ of thin cylindrical
shells of thickness $\Delta r$ and radius $r$ increases with the
cube of the optical frequency~\cite{hulst81}. In the relevant
range of cylinder radii, the Rayleigh-Gans model causes the
$\omega^{-3}$ dependence of $\ell$ in the Rayleigh-limit to be
reduced to $\omega^{-2.2}$ since~\cite{approxnote}
\begin{equation}
F(k_{\rm eff} r)\approx 0.488 (k_{\rm eff} r)^{-0.8}.
\label{eq:2D}
\end{equation}
For a hexagonal lattice of air cylinders in silicon with
$r/a=0.45$, typical for macroporous silicon
crystals~\cite{gruning95}, we find {$\ell \approx 40a$} for
frequencies near lowest order stop gaps, assuming a non-uniformity
$\delta R$ of 5\%. A much larger $\ell$ is required for integrated
circuit applications.

Many efforts currently focus on quantifying losses in 2D crystals
made from high index slabs on lower index cladding layers, for
which the nonuniformity $\delta R$ is around
5\%~\cite{soukoulis01,babacrete00}. Although the guided wave
profile normal to the slab is not incorporated in our model, we
believe that Eq.~(\ref{eq:2D}) yields a reasonable estimate of
scattering due to nonuniformity of the air holes in such
structures. Similar to 3D, applications of 2D structures in
photonic crystal integrated circuits require a formidable increase
in fabrication accuracies beyond the current
state-of-the-art~\cite{soukoulis01,noda03}. These scattering
losses add to currently widely studied out-of-plane scattering
that is intrinsic even to hypothetical perfectly fabricated 2D
crystal designs~\cite{benisty00}. In contrast to out-of-plane
loss, however, statistical variations cannot be reduced by design
optimization.

Scattering in photonic crystals opens opportunities to explore new
phenomena in multiple scattering of light~\cite{sheng}. Photonic
crystals allow unique control over fundamental aspects, such as
the transport velocity or anisotropies of light diffusion. A
fascinating application is the possibility to localize light,
which could serve to enhance non-linear
interactions~\cite{John87}. According to the usual Ioffe-Regel
criterion, Anderson localization occurs when the mean free path is
so strongly reduced that its product with the wave vector equals
one: $k \ell \simeq 1$. It has been proposed that in photonic
crystals this challenging criterion is relaxed to $k \ell \simeq
1/\sqrt{\rho_{s}}$, with $\rho_{s}$ the modification of the
photonic density of states (DOS) relative to free
space~\cite{Busch99}. Since the DOS is strongly reduced in
photonic gaps, localization of light may even be feasible with the
relatively long mean free paths predicted by our model.

We thank Allard Mosk, Peter Lodahl, Philip Russell, and Thomas
Krauss for stimulating discussions.  This work is part of the
research program of the ``Stichting voor Fundamenteel Onderzoek
der Materie (FOM),'' which is financially supported by the
``Nederlandse Organisatie voor Wetenschappelijk Onderzoek (NWO).''


\begin{thebibliography}{10}
\bibitem{transportnote}{Anisotropic scattering can be incorporated
in $F$ to obtain the transport mean free path~\cite{sheng}. The
difference is presently not relevant as anisotropy corrections are
modest for difference shells originating from displacements.}
\end{thebibliography}
\end{document}